\documentclass[epj]{webofc}
\usepackage[varg]{txfonts}   
\usepackage{graphicx,color} 
\usepackage{bm}       
\usepackage{amsmath}  
\usepackage{amssymb}

\usepackage[colorlinks]{hyperref}
\woctitle{21st International Conference on Few-Body Problems in Physics}
\newcommand{\bra}{\langle}
\newcommand{\ket}{\rangle}

\newcommand{\be}{\begin{equation}}
\newcommand{\ee}{\end{equation}}
\newcommand{\bea}{\begin{eqnarray}}
\newcommand{\eea}{\end{eqnarray}}
\newcommand{\beq}{\begin{equation}}
\newcommand{\eeq}{\end{equation}}
\newcommand{\beqa}{\begin{eqnarray}}
\newcommand{\eeqa}{\end{eqnarray}}

\begin{document}
\title{Understanding the proton radius puzzle:
Nuclear structure effects in light muonic atoms}
\author{Chen Ji\inst{1}\fnsep\thanks{\email{jichen@triumf.ca}} \and
        Oscar Javier Hernandez\inst{1,2}\fnsep\thanks{\email{javierh@triumf.ca}} \and
        Nir Nevo Dinur\inst{3}\fnsep\thanks{\email{nir.nevo@mail.huji.ac.il}} \and
        Sonia Bacca\inst{1,2}\fnsep\thanks{\email{bacca@triumf.ca}} \and
        Nir Barnea\inst{3}\fnsep\thanks{\email{nir@phys.huji.ac.il}}
}

\institute{TRIUMF, 4004 Wesbrook Mall, Vancouver, BC V6T 2A3, Canada 
\and
           Department of Physics and Astronomy, 
           University of Manitoba, Winnipeg, MB R3T 2N2, Canada
\and
           Racah Institute of Physics,
           The Hebrew University of Jerusalem, Jerusalem 91904, Israel
          }

\abstract{
We present calculations of nuclear structure effects to the Lamb shift in light muonic atoms. We adopt a modern {\it ab-initio} approach by combining state-of-the-art nuclear potentials with the hyperspherical harmonics method. Our calculations are instrumental to the determination of nuclear charge radii in the Lamb shift measurements, which will shed light on the proton radius puzzle.
}
\maketitle
\section{Introduction}
\label{intro}
The proton radius extracted from recent high-precision measurements of the Lamb shift (2S-2P atomic transitions) in muonic hydrogen atom $\mu$H~\cite{Pohl2010,Antognini2013} displays 7 standard deviations from results determined from $e$H spectroscopy or $e$-$p$ elastic scattering. This discrepancy raises intriguing questions about lepton universality and challenges theories of standard model physics. 
To resolve the proton radius puzzle, the CREMA collaboration at Paul Scherrer Institute started new measurements of the Lamb shifts in other light muonic atoms ($\mu$D, $\mu ^3\rm{He}^+$ and $\mu ^4\rm{He}^+$)~\cite{Antognini2011}. 
These measurements aim to extract the nuclear charge radii with extremely high accuracy, which will be benchmarked with current and future  spectroscopy measurements in electronic atoms~\cite{Parthey10,Rooij:2011,Cancio:2012} and electron-nucleus scattering data~\cite{MAMI:2012ed,Sick2014}.

\begin{figure}[htb]
\centerline{
\includegraphics[angle=0,scale=0.35]{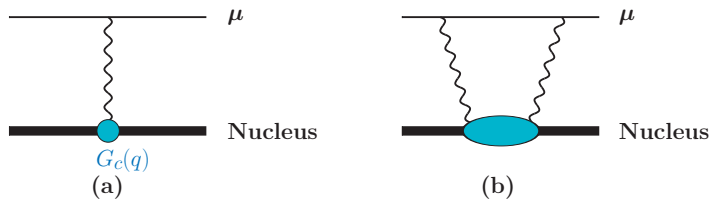}}
\caption{
The muon-nucleus one-photon (a) and two-photon (b) exchange processes: The bulb in (a) indicates the insertion of nuclear electric form factor. The bulb in (b) denotes the excitation of nucleus in the intermediate states between two photons.
} 
\label{fig:otpe} 
\end{figure}

The accuracy of the determined nuclear charge radius depends not only on the experimental precision of the Lamb shift, but also on the theoretical calculations. The theories relate
the Lamb shift $\delta_{LS}$ in a muonic atom/ion $\mu$X$^{(Z-1)+}$ to the charge radius  $\bra r^2_{ch} \ket$ of a nucleus $^Z_A$X by   
\begin{equation} \label{eq:E2s2p}
\delta_{LS} = \delta_{\rm QED} +  
                \mathcal{A}_{\rm OPE}\, \bra r^2_{ch} \ket +\delta_{\rm TPE}, 
\end{equation}
where $\delta_{\rm QED}$ is the contribution from the quantum-electrodynamics (QED) including the photon vacuum polarization, muon self energy, and relativistic recoil corrections. 
The other two terms in Eq.~\eqref{eq:E2s2p} are corrections due to nuclear structure.
The term linear to $\bra r^2_{ch} \ket$ is dominated by one-photon exchange between the muon and the nucleus (Fig.~\ref{fig:otpe}), where $\mathcal{A}_{\rm OPE} \approx m_r^3 (Z\alpha)^4/12$ with $m_r=m_{\mu} M_X /(m_{\mu}+M_X)$ denoting the reduced mass in the muon ($m_{\mu}$) and nucleus ($M_X$) system. 
$\delta_{\rm TPE}$ represents the two photon-exchange (TPE) contribution (Fig.~\ref{fig:otpe}), which is a combination of elastic ($\delta_{\rm Zem}$) and inelastic ($\delta_{\rm pol}$) parts. $\delta_{\rm Zem}$ is proportional to the third electric Zemach moment $\bra r_{ch}^3\ket_{(2)}$ by~\cite{Friar:1978wv}
\begin{equation}
\label{eq:zemach3}
\delta_{\rm Zem} = - m_r^4(Z\alpha)^5 \bra r^3_{ch}\ket_{(2)}/24~.
\end{equation}
The inelastic part $\delta_{pol}$ is the nuclear polarizability effect, where the nucleus in the muonic atom is virtually excited in the intermediate states by exchanging two photons with the muon.

The bottleneck in accurately extracting the nuclear charge radius in Lamb shift measurements is the theoretical uncertainty in determining $\delta_{\rm TPE}$, especially $\delta_{\rm pol}$. To determine $\bra r^2_{ch} \ket$ of $^3$He and $^4$He from measurements of their muonic ions to a $3\times10^{-4}$ accuracy, $\delta_{\rm pol}$, which strongly relies on our knowledge of nuclear structures and dynamics, has to be known within a 5\% accuracy~\cite{Antognini2011}. 

$\delta_{\rm pol}$ can be represented as a combination of energy-dependent sum rules of a series of nuclear response functions. One way to calculate $\delta_{\rm pol}$ is to use inputs from measurements of photo-absorption cross sections and quasielastic scattering data. However, the applications of this approach in $\mu$D~\cite{Carlson:2013xea} and $\mu\,^4$He$^+$~\cite{Bernabeu:1973uf, Rinker:1976en, Friar:1977cf} do not provide the accuracy desired by the Lamb shift measurements. The high-accuracy requirement can be achieved alternatively through calculations using state-of-the-art nuclear potentials. 
For example, the studies of nuclear structure effects in $\mu$D  
using Argonne $v_{18}$ (AV18) potentials and pionless effective field theory improved the accuracy to a $\sim 1\%$ level~\cite{Pachucki:2011xr, Pachucki:2015uga,Friar:2013rha}, which is one order of magnitude more accurate than previous methods.

We calculated the nuclear structure effects in $\mu$D using chiral effective field theory ($\chi$EFT) potentials~\cite{Hernandez2014}. Combining the modern inter-nucleon potentials with few-body methods, {\it i.e.}, effective interaction hyperspherical harmonics expansion (EIHH)~\cite{Barnea2000} and Lanczos sum rules (LSR)~\cite{Nevo2014}, we also calculated $\delta_{\rm TPE}$ in $\mu ^4\rm{He}^+$~\cite{Ji2013prl,Ji2013fbs}, $\mu ^3\rm{He}^+$ and $\mu^3\rm{H}$~\cite{Nevo2015}. 
By comparing the results using different potential models, we show that the uncertainty in $\delta_{\rm TPE}$ is dominated by nuclear physics.
Our results are key to the determination of $\bra r^2_{ch} \ket$ from future muonic atom measurements.

\section{Nuclear structure effects in $\mu$D}

$\delta_{\rm TPE}$ in $\mu$D was calculated by Pachucki~\cite{Pachucki:2011xr} with the implementation of AV18 nuclear potentials~\cite{Wiringa1995}. The dominant contribution is the sum rule of the electric-dipole response function, with corrections from the higher-multipole response function sum rules, relativistic and Coulomb distortion effects, {\it etc}. This calculation achieved a $\sim1\%$ accuracy by systematically analyzing neglected higher-order effects in atomic physics. Our recent work~\cite{Hernandez2014} provided an independent derivation of the formalism for $\delta_{\rm TPE}$, which is consistent with Ref.~\cite{Pachucki:2011xr}.\footnote{Small discrepancies between Refs.~\cite{Hernandez2014,Pachucki:2011xr} appear in the reduced-mass--dependent coefficients of higher-multipole contributions. These are resolved later by Pachucki {\it et al.} in Ref.~\cite{Pachucki:2015uga}, which agrees with Ref.~\cite{Hernandez2014}.} The later work by Pachucki {\it et al.}~\cite{Pachucki:2015uga} carefully 
analyzed the relativistic-recoil effects in $\delta_{\rm TPE}$, which indicate, overall, a few parts per mil correction. 

To fully analyze the uncertainty in nuclear physics, we used in Ref.~\cite{Hernandez2014} not only the AV18 potential but also $\chi$EFT with a series of parameterizations~\cite{Entem2003,Epelbaum2005}. We studied the order-by-order chiral convergence and the regulation-cutoff dependence. Our study shows a $0.6\%$ uncertainty from potential variations, which must be included in the total uncertainty of $\delta_{\rm TPE}$ in $\mu$D. With a further investigation, we include an additional correction $\delta^{hadr}_{\rm Zem}$ to our results in Ref.~\cite{Hernandez2014} to be consistent with Refs.~\cite{Friar:2013rha,Pachucki:2015uga}. This term is proportional to the proton Zemach moment, its contribution to $\delta_{\rm TPE}$ in $\mu$X$^{(Z-1)+}$, is related to the elastic TPE effects in muonic hydrogen $\delta_{\rm Zem}( \mu {\rm H})$ by
\begin{equation}
\label{eq:proton-Zem}
\delta^{hadr}_{\rm Zem} (\mu \textrm{X}) 
= - \left[Z m_r (\mu \textrm{X}) \right]^4\alpha^5 \bra r_p^3\ket_{(2)} /24
= \left[Z m_r (\mu \textrm{X}) / m_r(\mu H) \right]^4 \delta_{\rm Zem} (\mu \textrm{H})
\end{equation}
The hadronic polarizability in $\mu$X can be approximated from its effects in $\mu$H by
\begin{equation}
\label{eq:proton-pol}
\delta^{hadr}_{\rm pol} (\mu \textrm{X}) 
\approx   Z^3(Z+N)\left[m_r (\mu \textrm{X}) / m_r(\mu H) \right]^3 \delta^{hadr}_{\rm pol} (\mu \textrm{H})~.
\end{equation}
However, this estimate is based on the hadronic polarizability effect involving a single nucleon, but does not include the interference with nucleon-nucleon interactions. A more accurate calculation of $\delta^{hadr}_{\rm pol}$ in $\mu$D is done by Carlson {\it et al.} by using high-$Q^2$ part of the deuteron scattering data~\cite{Carlson:2013xea}. The combination of $\delta^{hadr}_{\rm Zem} (\mu \textrm{X})$ and $\delta^{hadr}_{\rm pol} (\mu \textrm{X})$ yields a total hadronic TPE contributions in $\mu$D to be $\delta_{\rm TPE}^{hadr}=-47(10)$ keV. 
Through collaborative communications among independent groups, $\delta_{\rm TPE}$ in $\mu$D is agreed to be $1.709\pm 0.020$ meV, which is adopted in a recent review by the CREMA collaboration~\cite{Krauth:2015nja} and will be used for extracting the deuteron charge radius in their Lamb shift measurement.

\section{Nuclear structure effects in $\mu^4$He$^+$, $\mu^3$He$^+$ and $\mu^3$H}
To extend the calculations to other light muonic atoms, we applied the EIHH method to expand the nuclear Hamiltonians and wave functions in the hyperspherical harmonic basis. This method is suitable for Hamiltonians with local (AV18) or non-local ($\chi$EFT) potentials. The energy-dependent sum rules of the response functions are obtained by implementing the LSR method~\cite{Nevo2014}.

We calculated the nuclear structure effects to the Lamb shift in $\mu^4$He$^+$, including the inelastic part of $\delta_{\rm TPE}$ due to nuclear excitations ($\delta^{nucl}_{\rm pol}$)~\cite{Ji2013prl} and the elastic part due to nuclear charge distributions ($\delta_{\rm Zem}$)~\cite{Ji2013fbs}. 
We have recently improved the treatment of nucleon size effects and we show in Table~\ref{tab-1} updated results of $\delta^{nucl}_{\rm pol}$ for $\mu$D and $\mu^4$He$^+$.\footnote{The $np$ correlated nucleon-size corrections to TPE were calculated in~\cite{Ji2013prl} with an inaccurate assumption that proton and neutron behave the same except for Coulomb. With a more rigorous calculation, we provide updated results.}
For the hadronic TPE effects, rigorous calculations using scattering data are not yet available. Therefore, we estimate such effects in $\mu^4$He$^+$ using Eqs.~\eqref{eq:proton-Zem} and \eqref{eq:proton-pol}. We find $\delta^{hadr}_{\rm pol} (\mu ^4\textrm{He}^+) \approx -0.34$ meV and $\delta^{hadr}_{\rm Zem} (\mu 
^4\textrm{He}^+)\approx -0.54(12)$ meV with the error dominated by the uncertainty in neutron polarizability.
$\delta_{\rm TPE}$ in $\mu$D and $\mu^4$He$^+$ are summarized in Table~\ref{tab-1}.

\begin{table}[htb]
\centering
\caption{Two-photon exchange contributions to the Lamb shift in light muonic atoms in units of meV.
}
\label{tab-1}  
\begin{tabular}{c | c c c c | c}

\hline
& $\delta_{\rm Zem}$  & $\delta^{nucl}_{\rm pol}$ & atomic error & $\delta^{hadr}_{\rm TPE}$ & $\delta_{\rm TPE}$ \\
\hline
$\mu$D  & -0.424(3)  & -1.245(5) & $\pm$0.018 & -0.047(10)& -1.716(20)\\
 
$\mu^4$He$^+$  & -6.29(28)  & -2.36(10) & $\pm$0.10  & -0.89(12) & -9.54(34)\\
\hline
\end{tabular}
\end{table}

We have recently extended our studies of TPE effects to $\mu^3$He$^+$ and $\mu^3$H. These muonic atoms of mirror isotopes, which contain unequal neutrons and protons, may provide additional information for the puzzle by probing the differences between muon-proton and muon-neutron interactions.
The calculations of $\delta_{\rm TPE}$ in $\mu^3$He$^+$ and $\mu^3$H will be detailed in an upcoming paper~\cite{Nevo2015}.

\section{Conclusion}

We illustrate {\it ab-initio} calculations of TPE contributions to the Lamb shift in light muonic atoms originated from nuclear structure. We provide the results for different muonic hydrogen and helium isotopes, which are important for determining the nuclear charge radii in Lamb shift measurements.

\begin{acknowledgement}
We thank Judith McGovern, Michael Birse, Aldo Antognini, Misha Gorchtein, and Krzysztof Pachucki for useful discussions. This work was supported in parts by the Natural Sciences
  and Engineering Research Council (NSERC), the National Research
  Council of Canada, the Israel Science Foundation (Grant number
  954/09), and the Pazy foundation.
\end{acknowledgement}

%
\bibliography{muA-TPE-proceedings.bib}
%
%

\end{document}